# **Title:** Spatio-Temporal Investigation of Brain-Wide Sequences

Ohad Felsenstein[1*], Moshe Abeles[1,2]

[1]Gonda Multidisciplinary Brain Research Center, Bar-Ilan University, Ramat-Gan, Israel.

[2]The Hebrew University of Jerusalem, Jerusalem, Israel.

**\*** Corresponding Author

E-mail: Ohad.Felsenstein@biu.ac.il (OF)

# Abstract

In "The Organization of Behavior" (Hebb, 1949), Hebb suggested that the propagation of activity between transiently grouped neurons plays an important role in behavior. Since then, multiple studies have provided evidence supporting Hebb's claim; however, most findings have been found locally in confined brain regions during unimodal tasks. Here we report on brain-wide behavioral-specific sequences in humans performing a multimodal task. To investigate the structure of these sequences, we used MEG to record brain activity in multiple brain regions simultaneously in participants performing a sensory-motor synchronization task. We detected local transient events corresponding to synchronously activating populations of pyramidal neurons and searched for their global organization as spatio-temporal patterns of activation sequences between distant neural populations. We focused our analysis on two types of spatio-temporal patterns: the *most frequently repeating patterns* and the *most discriminative patterns,* to concentrate on patterns with high relevancy to behavior. The findings revealed that global temporally precise sequences can be found, and that these sequences have partially stereotypical characteristics, both temporally and spatially, with consistent properties across subjects. By implementing a simplistic single-trial decoding approach, we found that brain-wide sequences have a temporal precision of 17-31 milliseconds, which resembles the temporal precision found locally in neural assemblies.

## Introduction

In 1949, Hebb's seminal book "The Organization of Behavior" presented a theory to explain behavior based on the most recent findings in the physiology of the nervous system (Hebb, 1949). One specific concept introduced in this book was the neural assembly hypothesis, which suggests that repeated stimulation may result in groups of synchronously activated neurons, or a group of neurons, firing in a prescribed sequence, which may transiently form a functional neural circuit. Since then, the term has been used to describe groups of synchronously activated neurons in a range of systems, with respect to their spatial, and temporal features, as reviewed by Gerstein (Gerstein et al., 1989).

According to Hebb, these cell assemblies play a fundamental role in the representation of information of distinct cognitive entities but are not sufficient for the generation of behavior. Hebb suggested the concept of "phase sequences" to describe the sequential activation of cell assemblies to bridge the gap to the generation of complex behavior. This sequential activation was based on a principle known as Hebb's rule which states that the repeated or persistent involvement of a presynaptic neuron in the firing of the postsynaptic neuron will result in a strengthening of the synapse (Hebb, 1949). Later studies have provided evidence to support these notions of cell assemblies and forms of phase sequences in terms of anatomy (Braitenberg and Schüz, 1991), theory (Buzsáki, 2010; Harris et al., 2003; Luczak et al., 2015), and in computational models (Abeles, 1982; Beggs and Plenz, 2003; Bienenstock, 1995; Dlesmann et al., 1999; Gerstein et al., 1989; Kumar et al., 2010).

The idea of stimulus-dependent chains of activations applies intuitively to the primary sensory areas. If the statistical properties of a repeating stimulus elicit a synchronous response in a primary sensory group of neurons, given Hebb's rule and spike-timing dependent plasticity (Bi and Poo, 1998; Markram et al., 1997), it is likely that synchronous activation will propagate to an adjacent group of neurons. In fact, stimulus-dependent sequences have been found in multiple studies on vision (Diana C Dima et al., 2018; Miller et al., 2014), the auditory (King et al., 2014; Luczak et al., 2009; Nelken et al., 2005), as well as the somatosensory (Derdikman et al., 2003; Devonshire et al., 2010), and olfactory systems (Cury and Uchida, 2010; Junek et al., 2010). However, one interpretation of the significance of phase sequences suggests that these sequences of activations could also underlie internal cognitive processes (Harris, 2005). This interpretation is supported by recent findings indicating that sequences emerge spontaneously during rest (Dragoi and Tonegawa, 2011; Luczak et al., 2009, 2007), decision making (Harvey et al., 2012; Jin et al., 2009), learning (Modi et al., 2014), recall (Pastalkova et al., 2008), and when generating bird songs (Hahnloser et al., 2002). Overall, these findings suggest that such temporally precise sequences may be a fundamental, intrinsic mechanism of the brain (Spreizer et al., 2019).

However, natural behavior is composed of more complex tasks requiring the involvement of multiple modalities and brain regions. While these temporally-precise sequences have mostly been reported in unimodal tasks, temporal precision could also occur in more complex multimodal behavior that involves coordination between multiple, possibly remote brain regions. The search for mechanisms allowing information to be routed, transmitted, and propagated between remote brain regions and subsystems has resulted in two primary transmission schemes, which are known as synchrony-based (Dlesmann et al., 1999; Kumar et al., 2010) and oscillation-based (Andre M Bastos et al., 2015; Fries, 2015). However, a recent study (Hahn et al., 2019) claimed that both could co-exist under a unified formulation and that they differ solely in terms of the speed of communication.

There are scant findings on the brain-wide propagation of activity in a sequence-like form, mainly due to the scarcity of imaging modalities that can measure brain-wide activity with high temporal resolution. Indirect evidence for brain-wide activity propagation with regards to phase-like sequences has been obtained by using a wide-field voltage-sensitive dye technique (VSDI). In these animal studies, stimuli that were presented visually (Han et al., 2008; Roland et al., 2006) or somatosensorily (Frostig et al., 2008; Mohajerani et al., 2013) elicited responses in sensory regions and later propagated as a wave to additional adjacent (Frostig et al., 2008) homological regions (Ferezou et al., 2007) or according to regional axonal projections (Mohajerani et al., 2013) and the rest of the cortex. These findings, in combination with reports of spontaneous activation sequences in the Hippocampus (Luczak et al., 2007), have led to the hypothesis that although activity may start in any cortical region it typically spreads to the whole cortex (Luczak et al., 2015).

Previous studies in our lab on humans performing multimodal tasks have identified intricate and highly precise brain-wide spatio-temporal patterns, but the number of times each pattern repeated was low (Tal and Abeles, 2016). We therefore directed our efforts to the search for possibly less precise interactions between pairs of sources and found that such interactions exist, and that these interactions conveyed a significant amount of information about the subjects' behavior (Felsenstein et al., 2019b, and Study II in this dissertation). Nevertheless, it remains unclear whether packet-like communication or sequences of brain-wide events can be found.

 The overarching goal of the current study was to investigate whether packet-like communication is present on the global scale of the cortex and cerebellar networks during a complex behavior task, and if so, the nature of its spatial and temporal characteristics. A *sensorimotor synchronization* (SMS) task was selected to elicit both sensory and internal cognitive processes. SMS involves the coordination of rhythmic movements to an external rhythm, which is usually tested in the lab through finger tapping experiments in synchrony with a metronome (Repp and Su, 2013). To perform SMS well, coordination between multiple systems, including the auditory, rhythm perception, timing, motor, feedback, and error correction systems must occur; for a review, see (Coull et al., 2011).

A Magnetoencephalogram (MEG) records brain activity throughout the cortex and cerebellum surfaces with high temporal resolution. The main contributors to the MEG signal are currents flowing along the apical dendrites of pyramidal neurons as generated by excitatory postsynaptic potentials (EPSP) close to the surface or inhibitory postsynaptic potentials (IPSP) in deeper layers (Creutzfeldt and Struppler, 1974; Lopes Da Silva, 2010). In order for these weak currents to be detected by the MEG, a local population of pyramidal neurons needs to operate synchronously, or a larger group of neurons must be in partial synchrony to generate a more substantial signal. This enables large cell assemblies to generate a transient signal that can be measured by the MEG.

In recent years a growing number of studies have reported behavior-modulated transient neocortical events in computer simulations, local field potentials(LFP), and MEG recordings (Jones, 2016; Sherman et al., 2016; Shin et al., 2017b; I. Tal and Abeles, 2018). Moreover, the rate (Shin et al., 2017a) and brain-wide density maps of these transient beta events (Abeles, 2014) can be used to successfully discriminate cognitive states in evoked and ongoing neuronal activity. While the neurophysiological basis of the generation of these transient events is in its initial stages (Law et al., 2019), a study on monkeys (I. Tal and Abeles, 2018) revealed an association between sudden increments in multi-unit activity and occurrences of these transient events in LFP signals. It is therefore possible that these transient beta events are at least partly a manifestation of cell-assembly activation in the MEG signal.

In the current study, we measured local transient events throughout the cortex and cerebellar surfaces to determine their global organization as spatio-temporal patterns. We focused our analysis on two types of behavior-related patterns, the *most frequently repeating patterns* and the *most discriminative patterns*. Based on these two pattern types, we investigated properties including the duration of these patterns, and the spatial and temporal relations between pattern sub-units and temporal precision. The findings revealed brain-wide behavioral-specific sequences in behaving humans that had particular spatio-temporal characteristics which were consistent across subjects. By implementing a simplistic single-trial decoding approach, we found that brain-wide sequences had a temporal precision of 17-31 milliseconds, which resembles the temporal precision found in local neural assemblies sequences.

## Methods

### Ethics statement

The internal ethics committee of Bar-Ilan University approved the protocol. All subjects signed an informed consent form before they participated in the study.

### Participants

Seven volunteers (5 female), right-handed (by self-report), and with no more than two years of formal musical training participated in the experiment. All participants were rewarded financially for their time and travel expenses.

### The behavioral task

In this study, we used a specific behavioral task as implemented in several studies by our group (Felsenstein et al., 2019b; I Tal and Abeles, 2013; Tal and Abeles, 2016) to investigate the spatio-temporal properties of patterns of activations elicited during rich multimodal behavior. The task is a combination of two commonly used paradigms. The first is derived from the field of sensory-motor synchronization (SMS) where subjects synchronize their movements to an external rhythm (Repp, 2005; Repp and Su, 2013), whereas the second is a variant of the oddball design (Näätänen, 2018) where two types of stimuli, one standard, one rare, are randomly presented to subjects. The task includes tapping along with an auditory stimulus that forms a musical meter (Tal and Abeles, 2016). We expected that a task that incorporates sensory, motor, attention, working memory, and other cognitive processes would elicit rich activity across a network of information-processing brain regions.

*Fig* illustrates the task. Participants listened to a sequence of drum beats that generated a particular meter (e.g., 3/4), and tapped along with the beat using their index finger for the accented (primary) beats and the middle finger for the unaccented (secondary) beats. At random points in time (marked with dotted lines in the figure), the meter changed, and the subject had to adjust the tapping accordingly (Tal and Abeles, 2016). These points of change in meter created two conditions: (1) *Before-change*: the last 1.25 seconds with synchronous taps (see two examples marked in blue in *Fig* ); (2) *After-change*: the first 1.25 seconds after ten milliseconds from the auditory onset (see examples marked in red). For a short video with an illustration of the task, see (Felsenstein and Chechik, 2018); note that the trial definition was slightly different in the video.

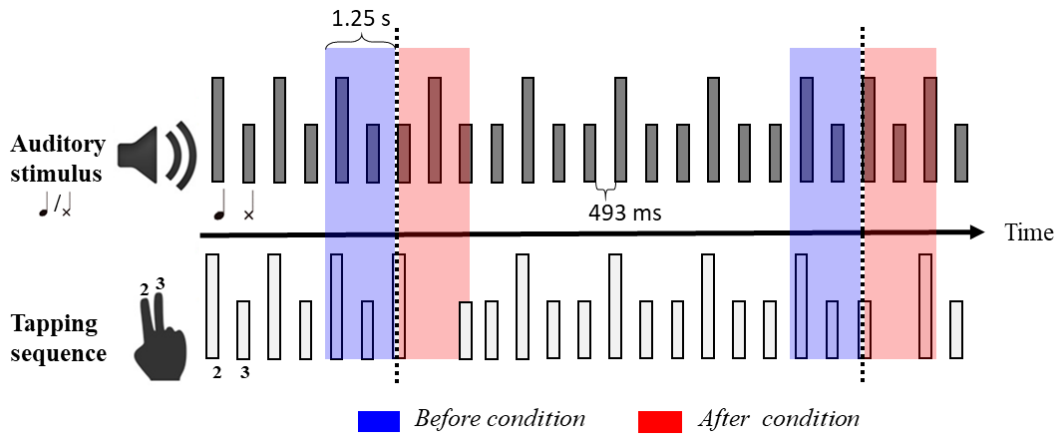

**Fig 1. The behavioral task.** The auditory stimulus consisted of a sequence of drum beats: Tall bars indicate primary beats, and short bars indicate secondary beats. **Bottom:** The tapping sequence. Tall bars indicate tapping with the 2nd finger, and short bars indicate tapping with the 3rd finger. The dotted line indicates changes in musical meter. Blue shading indicates the 1.25 seconds defined as the "before" condition, immediately preceding a change in meter. Red shading marks the 1.25 seconds defined as the "after" condition, starting 10 ms after the auditory stimulus onset indicating meter change. Note that for periodic stimuli, tapping precedes the sound.

Specifically, subjects listened to a sequence of drum beats composed of two different auditory stimuli with a fixed beat-beat interval of 0.493s. The two sounds formed one of two structures (musical meters), where a primary beat was followed by *one* secondary beat (double meter), or where the primary beat was followed by *two* secondary beats (triple meter). During the experiment, the structure changed at random after at least five consecutive repetitions of the same structure. Since the probability of a pattern repeating with high accuracy was low, we expanded the definition of a trial to the 1.25 seconds before or after the change in musical meter.

The definition of trials before and after the point of change in meter yielded an identical number of trials for the two conditions, with an average across subjects of ~88 trials per condition. The actual number of trials per condition varied across participants from 82 to 93, depending on the cleaning procedure for individual participants. Importantly, trials within each condition contained an equal number of trials representing the two types of congruency or incongruency; namely, hearing a primary or secondary beat and tapping with the middle or index finger respectively for the *after-change* condition and similarly for the *before-change* condition. By preserving this balance, the analysis could center on higher cognitive processes rather than merely the discrimination between sounds or motor actions.

From an oddball design point of view, the subjects listened to a regular auditory stimulus forming one of two types of structure: a 2/4 meter (a sequence of an accented beat followed by an unaccented beat) and a 3/4 meter (an accented beat followed by two unaccented beats). Consecutive repetitions of the same musical meter formed the local standard, and the random change in musical meter following these repetitions created the local deviant. On a smaller scale, the structure of accented and unaccented beats that formed the musical meter could be viewed as a global regularity of XY or XYY (Bekinschtein et al., 2009). In this case, the matching beats in the global standard case constituted the *before-change* condition, and the deviant beat formed the *after-change* condition.

## MEG data acquisition

This article is a re-analysis of the data collected in (Tal and Abeles, 2016). Here, we analyze a subset of the subjects who were recorded at a high sampling frequency of 1017.25Hz. Below we briefly describe the data acquisition process; for more details, see (Tal and Abeles, 2016). A whole-head helmet-shaped biomagnetometer (4D-Neuroimaging, San Diego) with a sensor array consisting of 248 superconducting magnetometers was used to acquire the MEG data. Subjects were placed in the supine position so that changes in head position relative to sensors were minimal. Five coils, attached to the scalp, provided information on the head position relative to the sensor array before and after the measurement. Coil positions were determined based on external anatomical landmarks (left preauricular, right preauricular, and nasion). The head shape and coil positions were digitized using a Pollhemus FASTTRAK digitizer. MEG signals were band-pass filtered online at 0.1-400Hz.

### MRI data acquisition

Magnetic Resonance Imaging (MRI) data were collected using a 3T scanner (Signa Excite, General Electric Medical Systems, Milwaukee, WI) located at the Tel Aviv Sourasky Medical Center. High-resolution T1 anatomical images were acquired for each participant using a 3D fast spoiled gradient-recalled echo sequence (FSPGR; 150± 12 1-mm thick axial slices, covering the entire cerebrum; voxel size: 1×1×1 mm). For more details, see (Blecher et al., 2016).

### MEG data preprocessing

The MEG recordings were cleaned using the procedure described in (I Tal and Abeles, 2013) that included the power-line frequency, heartbeat artifacts, and 24Hz building vibration artifacts. Power line artifacts were eliminated using a trigger time-locked to the line oscillations. We removed the heartbeat artifact, for each channel, by computing the mean heartbeat artifact around the QRS peaks of the heartbeat and subtracting it from the MEG channel. To reduce the outside vibrations artifact, gantry vibrations were recorded simultaneously during the experiment and were used to clean the signal in the frequency domain. Eye movements and eye blink segments were detected using a spatial ICA algorithm implemented in the FieldTrip® Matlab software toolbox for MEG analysis (Oostenveld et al., 2011). Trial pairs defined by segments spanning ($t_0$-1.25 s to $t_0$) and ($t_0$+0.1 s to $t_0$+1.26 s), where $t_0$ was the time of meter change which contained eye movements or eye blinks were visually inspected and discarded as a whole. For an in-depth evaluation and comparison of the cleaning procedure used here to other cleaning procedures, see (I Tal and Abeles, 2013).

In common practice, the source reconstruction of brain activity is based on segmenting the entire brain into a few thousand voxels and estimating the signal in each of these voxels (usually measuring 5 x 5 x 5 mm). However, using this method was unsuitable for the analysis described here since it would lead to an inflated dimensionality of the data, which would increase computational complexity and could impair the decoding accuracies. We therefore reduced the input dimensionality by estimating signals at 131 equidistant *locations-of-interest* (LOIs) on the dorsolateral aspect of each individual subject's cortex and cerebellar surfaces. The procedure for generating a set of locations for each subject which was consistent across subjects was implemented using the MMVT (Felsenstein et al., 2019a, and study I of this dissertation) tool as described below.

We started by computing the equidistant coordinates for the template brain of Colin 27 (Holmes et al., 1998). First, we created a geodesic polyhedral covering the cortex and cerebellar surfaces and projected the vertices to the closest point on the brain surface, primarily on the gyri. The purpose was to avoid selecting LOIs at the depth of the sulci and to avoid possible problems due to imperfect co-registration between the digitized head shape of the subject and the anatomical MRI. We discarded locations not situated on the dorsolateral aspect of the brain because of the substantial distance from the MEG sensors, yielding a total number of 131 LOIs. *Fig 2a* presents the geodesic polyhedral (shown in cyan) and the projected LOIs (shown in blue) over the Colin 27 template brain surfaces. The mean distance between neighboring LOIs was 2.35 cm (SD 0.31).

Next, we used Freesurfer (Anders M. Dale et al., 1999) to compute the transformation matrix from each individual subject to the template brain and to create a three-dimensional reconstruction of the cortex and cerebellar surfaces based on the subject's T1 scan. Using the transformation matrix of each individual subject, we computed the LOI transformed location and projected the transformed location to the individual surface to obtain the final LOI location. The full procedure yielded a set of locations for each subject that was consistent across subjects.

### Source localization

This study was designed to inspect spatio-temporal patterns occurring in close temporal proximity (up to tens of milliseconds) and multiple LOIs situated in various brain regions. One potential pitfall was that the patterns and spurious connectivity between the two projected MEG time series could be a crosstalk artifact originating from the source localization procedure. The dependence between signals reconstructed at spatially-separated brain locations is known to be a limitation of source localization (e.g., Gross et al., 2001), and is explained by the ill-posed inverse problem (RamÃ­-rez, 2008). To lessen the problem of dependency between sources, we applied the Octahedron method (Lots Shapira et al., 2013), which was shown to reduce the correlations between adjacent sources.

The amplitudes of the current dipoles at the 131 LOIs were evaluated using the following procedure. We constructed a cube measuring 2x2x2 cm around each LOI and rotated the cube in 27 different orientations around the cube's cardinal axes. For each rotation, we used the simultaneous N-source reconstruction based on simultaneous synthetic aperture magnetometry (Robinson and Vrba, 1999) on the 1+8 points. *Fig 2a* presents the cube (in black), four of the eight points (in green), and the central point (in red). The reconstruction was based on the calculated covariance matrix on the MEG data filtered at 15–30 Hz. The rotation with the minimal correlation between the center point signal and the eight surrounding points was chosen. We refer to the signal in the center point evaluated by this method as the "cortical current dipole" amplitude (CCD).

This procedure yielded 131 CCDs, each spanning the entire recording period. *Fig 2b* and *Fig 2c* depict the signal in three neighboring positions. *Fig 2b* depicts the signals recorded in three neighboring MEG sensors, whereas *Fig 2c* shows the signal in three neighboring LOIs as obtained following the procedure described here. Note that as expected, the correlation between neighboring points was reduced significantly. For a detailed description of this method with evaluations of its contribution to reducing the correlation between neighboring points, see (Lots Shapira et al., 2013).

*Fig 2d* depicts the evoked CCDs for the two behavioral conditions measured in Subject 1. Each CCD trace is the signal in a single LOI filtered at 1-60Hz, averaged around the beat, and normalized to have 0 mean and unit variance. The blue and red traces show CCDs in the *before-change* (congruent) and *after-change* (incongruent) tapping conditions.

As expected in an oddball design, responses to the unexpected beat in the *after-change* condition exhibited well-known components such as mismatch negativity (MMN) for both the early event-related component at 150-250 ms and the later component around 300 ms from stimulus onset.

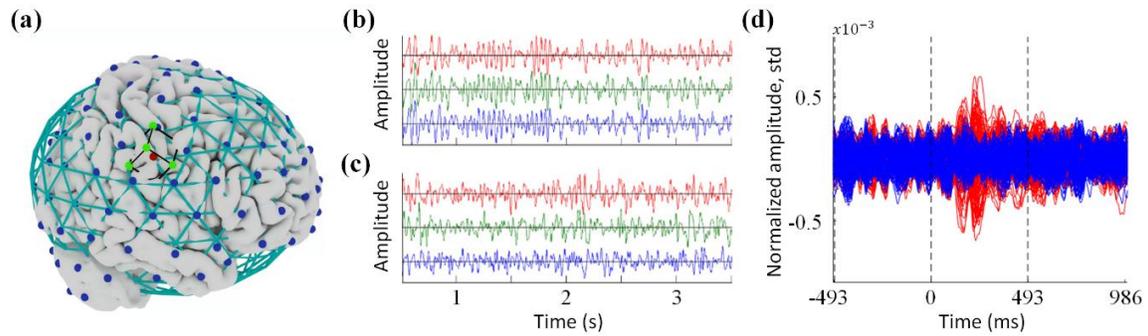

Fig 2. Data preprocessing, from the MEG signal to the source localized CCDs. **(a)** LOI distribution overlaying a template brain. The vertices and edges of the initial geodesic polyhedral, which are outside of the brain, appear in cyan. The blue points show a subset of the 131 projected LOIs. The red point shows a single LOI. The green points are four of the eight surrounding points in the shape of a cube. **(b)** MEG signals in three adjacent MEG channels. **(c)** CCD signals in three neighboring LOIs. **(d)** CCDs for evoked responses in Subject 1. Superposed 131 CCDs for the two conditions. The blue and red traces show amplitudes around the congruent and incongruent beat, respectively.

Mini-ER

Since the goal of this study was to investigate the spatial and temporal characteristics of possibly precisely-timed brain-wide activations, we used a point-process representation of the MEG signal as computed by (Abeles, 2014). The transformation of the analog CCD signals to a point process was done by detecting a sharp activity transient of half of a beta cycle as reported in multiple brain regions and multiple modalities such as MEG, EEG, and LFP (Jones, 2016; Sherman et al., 2016; I. Tal and Abeles, 2018). A recent study on monkeys by (I. Tal and Abeles, 2018) revealed an association between a sudden increment in multi-unit activity and the LFP signals that contained these transient events, which was termed the mini-evoked response (mini-ER). The rate (Shin et al., 2017a) and density maps of transient beta events across the brain (Abeles, 2014) were shown to be informative of the behavioral condition and could discriminate cognitive states in evoked and ongoing neuronal activity successfully.

Here we detected transient events and transformed the analog signals into a parallel point-process by applying the following approach. First, we obtained an analog template collecting a few tens of occurrences of such transients manually, and ran a principal component (PC) analysis which showed that a single PC accounted for more than 95% of the signal variance. We therefore used the first PC. We added a constant segment before its beginning to avoid finding transients sequentially due to periodic oscillations. By adding this constant, we increased the probability of detecting the first transient in every sequence. The basic template obtained from this process is shown in *Fig 3a*.

Next, the automatic detection algorithm described in (Abeles and Goldstein, 1977) was applied to detect occurrences of the transients of interest. The fit (S) of the template to the signal was calculated by convoluting the template (T) with the normalized signal (V). The residual (E) was computed by E=V-ST. Finally, the SNR was computed as $\frac{S^2}{\|E\|^2}$. This SNR showed sharp peaks, and those that exceeded a pre-set threshold were considered points of optimal match to the template. Next, a fine-tuned template was computed for each LOI separately by averaging the signal with respect to signal sign around these detections in the specific LOI. These LOIs' specific templates were used for a second pass of the same process to obtain the final detection of transients. The final detection of transients was done similarly by detecting peaks in the SNR, but we split the detected events according to the signal polarity for each LOI. *Fig 3b* presents an example of the analog signal and the positive and negative detected events. The separation between positive and negative events was used to discriminate between groups of neurons on the other bank of a sulcus which would produce a negative template. We set a threshold for each LOI's SNR separately to yield a rate of 5 transients per LOI per second, regardless of the polarity of events. Finally, for purposes of presentation, we created two separate LOIs for each LOI as depicted in *Fig 3c* by the vertical and horizontal tensors such that all positive events found in a specific position were associated with the vertical LOI and the negative events were associated with the horizontal LOI in the same position. The detected events were defined as the mini-evoked-responses (mini-ER). *Fig 3d* shows a raster-plot presenting the detected events found in 262 rows corresponding to the 131 LOIs as a function of the mini-ER event polarity. Two consecutive trials of before-change and after-change conditions are shown.

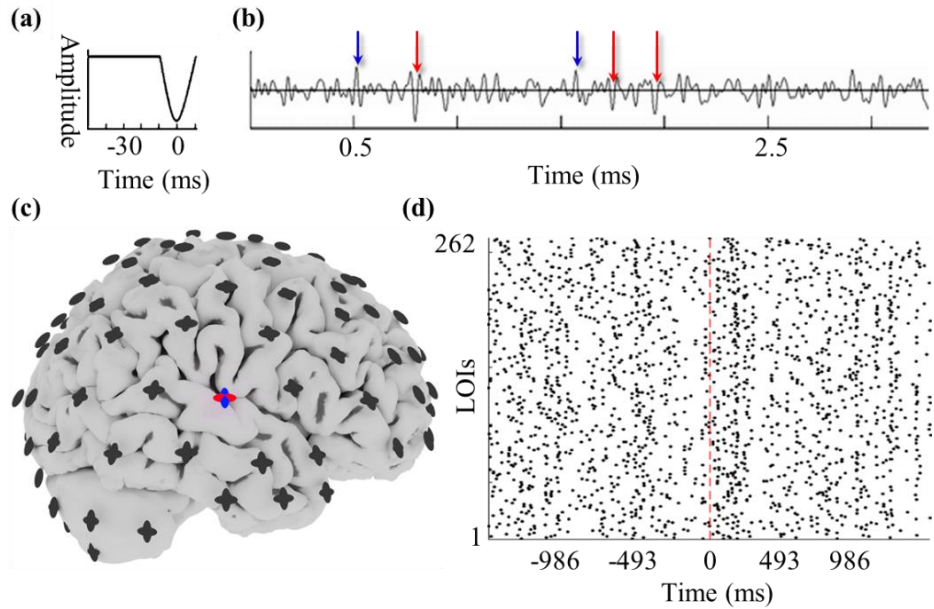

Fig 3. Data preprocessing, from CCDs to point-process. (a) The wavelet used for the detection of the mini-ER in the analog CCDs. (b) An illustration of a CCD time course and five points detected for mini-ERs. The blue and red arrows mark the points in time where the positive and negative mini-ERs were found, respectively. (c). Final LOI distribution. Each location of the original 131 LOIs was split into two channels. For example, the negative events found in the signal shown in (b) were assigned to the horizontal red tensor, and the positive events were assigned to the vertical blue tensor. (d) Brain-wide activity pattern: the activity recorded during a single trial in each condition across all 262 LOIs for Subject 2. Each row corresponds to a single LOI (shown as tensors in (c)). The dashed red line shows time zero; i.e., the time of the first auditory stimulus that indicates a change in meter. Each point denotes the detection of a mini-ER event.

Pattern detection procedure

In the following section, we describe how the wide activation patterns were extracted from the parallel point-process detailed above. The first step consisted of exhaustively extracting all patterns composed of three events with a maximal time-length of T milliseconds occurring during trials for individual subjects. Each pattern was defined by the LOIs and the time delays between events.

The next step was designed to find patterns that repeated multiple times with a defined temporal precision. Intuitively, greater temporal precision is likely to be associated with fewer repetitions. We implemented all the analyses, except the neural decoding experiments, at a temporal resolution of $\pm 8\ ms$. Here, our assumption of a temporal resolution of $\pm \Delta\ ms$ indicates a tolerance of $\Delta$ milliseconds in moving an event from its original time of occurrence. However, manipulating the signal raises issues such as how temporal shifts should be done, and how to determine which pattern should be defined as a prototype for the grouping of patterns.

For example, a long list of delays can be split into groups in many ways within a given delta. For example, assume the delays are 1, 2, … 10, and that the temporal resolution is 3. This list can be broken down into two groups <1, 2,…5> and <6,7,…10> or into one large group <1, 2,…8> while ignoring 9 and 10. The algorithm we implemented favors the largest group even if it meant that some data were lost.

To do so, we created a method for achieving a principled grouping of patterns based on a simple clustering method. However, in order to use a clustering method, we first needed to represent the data in a coordinate system. Since in the method described here, we did not attempt to allow for imprecision in space, we conducted the next procedure separately for each set of three LOIs. Our embedding of patterns composed of three delays between events was based on the snowflake representation described in (Abeles, 1983; Perkel et al., 1975). This representation capitalizes on the fact that although we can compute three delays for a given pattern, the number of degrees of freedom is only equal to two. By using this representation, we can obtain the <x, y> coordinates for every combination of delays between events within a predefined maximum length. *Fig 4a* depicts this representation, as presented in the original paper (Abeles, 1983).

Next, we convolved the coordinates matrix with a square window of size $(2*\Delta + 1) \times (2*\Delta + 1)$. Here convolution was used to estimate the number of clusters by counting the peaks in the convolved signal. Peaks within the distance required by the temporal resolution were combined and counted as one peak. Points with distances exceeding $t$ from peaks were discarded, and the remaining points and the estimated number of clusters were fed into a k-means algorithm. An $\ell_1$ norm was used for computing distances in the k-means algorithm to reduce run-time complexity.

Finally, a final run over the data was conducted to assure that all the clusters complied with the temporal resolution requirements, and clusters containing fewer than four points were discarded.

*Fig 4b* depicts an example of the 342 times a pattern composed of events in LOIs 54, 103, and 209 emerged. Each point corresponds to a single detection of a pattern, and the colored dense areas show the clusters found using the procedure described above. Since the maximal delay within each pattern was limited, all occurrences appeared inside a hexagon shape, as shown in *Fig 4a*. Note that the <x, y> coordinates used are orthogonal whereas the coordinates of the snowflake plot are not. Hence, at times the jitter within each group exceeded the temporal resolution.

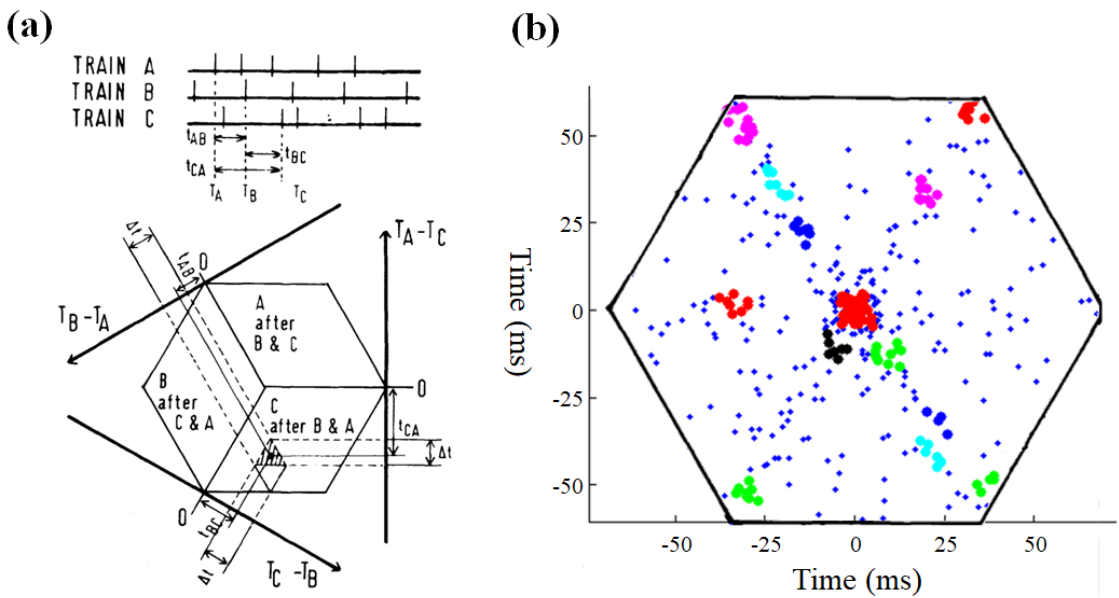

Fig 4. Data preprocessing, from point-process to spatio-temporal patterns. (a) The coordinate system of the three-cell correlation. Upper part: Three arbitrary event trains. The three events occurring at TA, TB, and TC are characterized by three intervals: tAB = TB - TA, tBC = TC - TB, and tCA = TA - TC. Lower part: The three intervals are represented by a unique point in the coordinate system. The triangular bin used to quantify the histogram is shown around that point. (b) The triplet <54, 103, 209> repeated 342 times. The three delays in all these 342 triplets are shown using a snowflake plot. Most data are on the three main axes, thus indicating that one of the delays is close to 0. Points in dense areas are colored to show the clusters with a temporal resolution of $\pm 5m$. In this illustration, the minimal number of points required for defining a cluster was five.

The process described here yielded a list of centroids representing the clusters, and generated the times in the experiment in which a pattern from that cluster occurred. The analysis of pattern detection and clustering requires immense computational resources and time. Thus, we added a few additional steps to the implementation to increase the rate of convergence and coding efficiency. The source code used in this analysis is available online (O Felsenstein and Abeles, 2020).

## Significance estimation for the spatial relations analysis

The significance of the spatial relations analysis was estimated as follows. Since the rate of events in each position was predefined and equivalent across positions, a single pattern was sampled randomly by drawing three LOIs from a uniform distribution over the 262 optional LOIs. A single permutation was calculated by repeating this sampling K times (between $10^2$ and $10^6$) and calculating the two-dimensional histogram for the permutation. In total, $10^5$ permutations were sampled for each of the five optional K values. The p-value for each bin was calculated by the normalized rank of the real data bin value in the corresponding bin permutation distribution.

## Significance estimation for the temporal relations analysis

The following procedure was used to estimate the significance of the temporal relations analysis. The null hypothesis was that the time difference between the first and second events and the second and third events in a pattern was random, but the summation of the two time differences could not exceed a maximal time span of 80ms. To test this null hypothesis, every single pattern was sampled randomly by drawing two time difference values from the empirical time difference distribution of all patterns and assuring that its summation did not exceed 80ms. Repeating this procedure to sample K triplets (for the range of $10^2$ to $10^6$) resulted in a single permutation value. In total, $10^5$ permutations were sampled for each of the five optional K values. The p-value for each bin was calculated by the normalized rank of the real data bin value in the corresponding bin permutation distribution.

## Decoding pipeline

In our attempt to investigate the temporal resolution of the patterns, we faced the challenge of the absence of a proper baseline for temporal resolution. In the end we used a previously applied method (Felsenstein et al., 2019b) which consists of comparing the decoding results on a range of candidate values. The decoder results are used to compare the likelihood of a specific parameter in comparisons to other parameters. Specifically, we investigated the temporal resolution of the patterns by decoding the behavioral condition based on the simple ensemble classifier described below.

We started by clustering the patterns according to the specific temporal resolution described in detail in the previous section. We then represented the data as a matrix $X \in \mathbb{N}^{\#clusters \times \#trials}$ where each entry was an indicator of whether cluster $i$ was found in trial $j$. We split the data into train and test sets where the test set was composed of two trials, one for each condition.

Our naive decoding approach was based on three steps:

### Step 1: Computing the discrimination score for each pattern.

We started by computing for each cluster $i$ how many times, in total, it appeared in training trials of behavior 1 ($N1_i$) and how many times it appeared in behavior 2 ($N2_i$). Next, we used the binomial cumulative distribution function to calculate, for each cluster, the probability of finding N1 or more occurrences (successes) of N1+N2 experiments under the assumption that the clusters were independent of the behavioral condition. Similarly, we calculated the probability of finding N2 or more occurrences of N1+N2 experiments and selected the minimal probability between the two. For each cluster we defined the discrimination score as 1- *p-value*, such that the more the cluster was discriminative, the more the discriminative score approached 1.

Step 2: Defining cluster selectivity.

In the second step, we constructed a selectivity vector $w \in \{-1,0,1\}^{\#clusters \times 1}$ that defined the cluster's behavioral condition selectivity. We set $w_i$ to be zero for all clusters with a discrimination score below a predefined threshold $\alpha$. For clusters with a discrimination score above $\alpha$ we set

$$w_i = \begin{cases} -1, N1_i > N2_i \\ 1, N1_i < N2_i \end{cases}.$$ Note that in the *before-change* condition subjects responded routinely, but in the *after-change* condition the subjects had to adjust to the change of meter. The inherent variability of *after-change* trials caused an imbalance in selective clusters for the two conditions. We therefore randomly zeroed the weights of k *after-change* selective clusters to equalize the number of selective clusters in the two conditions.

Step 3: Inference.

At inference time, we classified a test trial $j$ by the plurality vote of selective clusters by computing $y = sign\left(w^T x_j^{test}\right)$.

Evaluation

We used a stratified variant of the leave-one-out cross-validation procedure to evaluate decoder performance.

The development of the decoding method and the hyper-parameters search was conducted on the data of a single subject that is not presented in the current work, to reduce the likelihood of an over-optimistic evaluation.

# Results

In the following section, we describe several spatial and temporal properties of spatio-temporal patterns. Specifically, we focus on studying sequences composed of three events with a precise temporal and spatial structure that was repeated at least four times. A sequence fulfilling these requirements is referred to as a *pattern,* and the sources in it as a *triplet*. While the number of patterns that met the requirements was immense (on the order of $10^8$ to $10^9$), there was no guarantee that these patterns were relevant to the behavior of the subjects during the task. We therefore decided to narrow down the number of patterns by introducing one of two additional pre-conditions.

The first pre-condition required a higher number of exact sequence matches to define it as a pattern. By setting a higher threshold for the detection of a pattern, we increased the signal to noise ratio, and expected that patterns that repeated many times would correspond to features that were relevant to the two behavioral conditions, such as the auditory, motor, and attention systems. The second pre-condition was the selectivity of a pattern to the behavioral condition. To find these discriminative patterns, we modelled the probability that a pattern would repeat many times during one behavior but only a few times during the other behavior. For more details on the discrimination score, see the decoding pipeline section in *Methods*. In the following section, we used the scores obtained by enforcing each of these two pre-conditions to sort the patterns and inspect the characteristics of the top-ranked patterns.

We started by inspecting the spatial characteristics of the top-ranked patterns. From here onwards, we compared two pattern types, the *most frequently repeating patterns* (MRP) and the *most discriminative patterns* (MDP). We first focused on the spatial relationship between LOIs within the top-ranked patterns at the individual subject level. To do so, we computed a density histogram corresponding to the spatial distance between sources in a pattern. This type of analysis can indicate whether top-ranked patterns, either due to the number of repetitions or discrimination, exhibit a local or global form of spatial distribution. We computed ten two-dimensional histograms for each subject, one for each cutoff value between the top $10^2$ and $10^6$ and pattern type. Next, we computed the *p-value* for each entry by a permutation test, as described in the *Methods* section. Since the results for each subject were similar, *Fig 5* shows the median p-values across subjects obtained for each bin in the two-dimensional histograms.

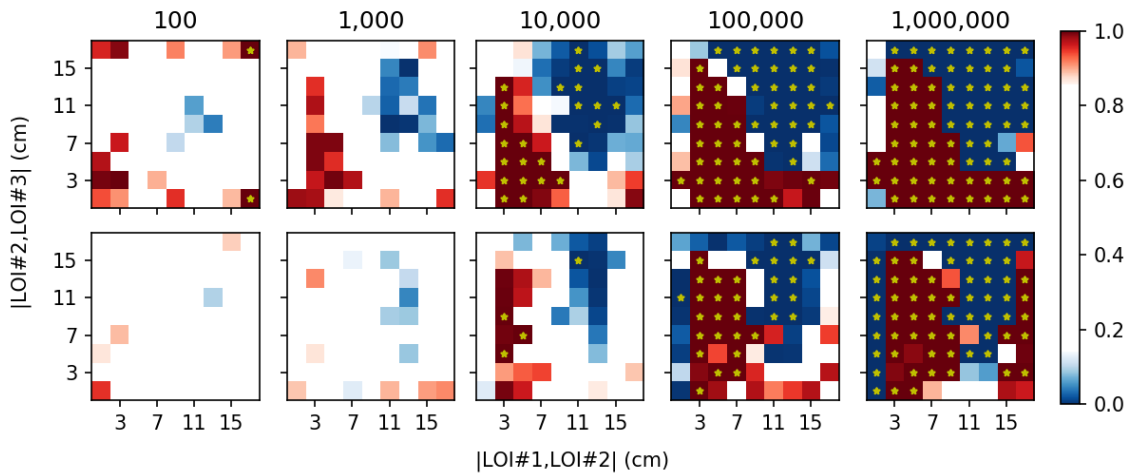

**Fig 5. Spatial relations of top patterns.** Each panel shows the median p-values across subjects for the distances between the first and second sources (shown on the x-axis) and the second and third sources (shown on the y-axis) for each pattern. The top and bottom rows show the histograms for the MRP and MDP pre-conditions, respectively. Each column shows the results for a different cutoff value yielding the number of patterns analyzed. Color depicts the median $p-value$ across subjects obtained by $10^5$ permutation tests. Red and blue indicate a higher or lower number of counts in the data respectively compared to the permutation counts; the yellow asterisks depict bins with a

$p-value < 0.001$ for both tails.

This analysis revealed the typical spatial relations between sources contributing to the same pattern for the top MRP and top MDP. A non-random formation emerged in both the MRP and MDP histograms, since clusters with significantly greater and lesser patterns emerged. These clusters revealed a semi-local formation where patterns composed of at least two LOIs at a short distance of up to 8 centimeters tended to occur more frequently than by chance. This formation was preserved across cutoff values and became more pronounced as the cutoff values increased, and the sample size became larger.

There was a general similarity between the spatial organization of the MRPs and MDPs that presented an asymmetry in patterns as the distribution of distances between the first two LOIs in the pattern and the last two LOIs of the pattern differed. Although the MRPs and MDPs showed a general similarity, there was a difference between the two pattern types mostly for the shortest distances on both LOI pairs and greater maximal distances between the first and second LOIs in the MDP. These differences could have originated from the different cognitive processes and brain regions activated during tapping and listening to an auditory stimulus, as well as during active identification of incongruency or attempting to resync with the new rhythm.

The next step was inspect the spatial organization of the top patterns on the individual and group levels. For the individual-level analysis, we extracted and visualized the top 15 patterns for each pattern type. *Fig 6a* and *Fig 6c* depict the patterns and LOIs obtained for Subject 1 over his cortex and cerebellar surfaces. The pink and green connections in *Fig 6a* and *Fig 6c* highlight specific patterns. For example, the pink triplet is composed of LOIs in the left supplementary motor area, left central sulcus and right superior temporal sulcus, and the green triplet is composed of LOIs in the left cerebellum, left precentral gyrus and the right Pars-triangularis. The LOI colors show the number of patterns involving an LOI out of the $10^4$ top-ranked patterns for that individual subject.

We repeated the procedure for all subjects and looked for similarities across subjects, but since only 15 patterns were present per subject and condition, finding similarities in such a small set would have been highly unlikely. Instead, we analyzed the overlap between subjects as follows. For each subject, we grouped all patterns composed of the same sources and in the same order of activation. This grouping procedure yielded $262^3$ optional triplets (ordered sampling with replacement), and since the LOI locations and naming were consistent across subjects, we could compute the score for each of the pre-conditions by computing the median for each triplet. *Fig 6b* and *Fig 6d* depict the top ten ranked triplets across subjects overlaid on Subject 1's cortex and cerebellar surfaces for consistency. The LOI colors depict LOI participation in the $10^4$ top-ranked triplets for the group of subjects. The pink triplet is composed of LOIs in the left cerebellum, left central sulcus, and left the supplementary motor area, and the green triplet is composed of LOIs in the left precentral gyrus, left cerebellum and right pars opercularis. The right side of each panel in *Fig 6* is an interactive 3D visualization of the connections and LOIs inclusion measures. The interactive panels can be manipulated by clicking on the panel and once the manipulation sign appears, by hovering the cursor while pressing the left button (panel interactivity is enabled in Adobe Acrobat 2017 or newer).

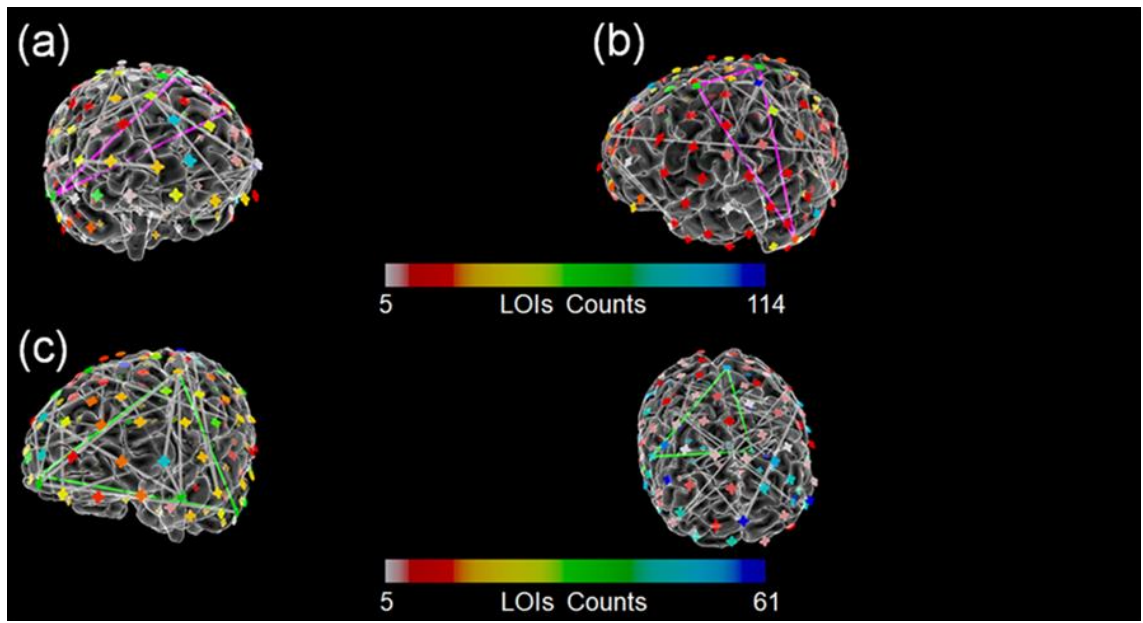

Fig 6. Spatial organization of top patterns. LOIs and connections outline the fifteen (a) MRPs and (c) MDPs. (b) The top ten most frequently repeating sets of ordered LOIs across subjects and (d) the most discriminative sets. Each panel shows the patterns or triplets overlaying the cortex and cerebellar surfaces of Subject 1 on the left, and the interactive 3D visualization of the same patterns or triplets, and LOI on the right. LOI colors depict the number of patterns out of the top $10^4$ involving LOI according to the matching color bar. Exemplary patterns are presented by colored connections as described in the text.

Visual inspection of these results showed that the top-ranked patterns were not confined to specific brain regions. Instead, the top-ranking patterns involved synchronous activation in various brain regions, and these patterns did not show a specific local distribution. The expected involvement of the left motor and somatosensory cortex, cerebellum, and temporal areas was found. In both MRPs as well as the MDPs, the inferior frontal gyrus (IFG) was involved bilaterally. The left IFG may have been activated as a result of inner speech used by the subjects to follow the meter composed of two tones. The right IFG, which has been linked to syntactic violations (Kaplan et al., 2010), was only found for the top MDPs, which is consistent with the syntactic violation caused by the discrete beat that violates the larger anticipated structure (Fiveash et al., 2018).

The group analysis revealed for the most frequently repeating triplets that LOIs involved in the highest number of triplets were situated in the superior frontal gyrus bilaterally, the right middle temporal gyrus, the central sulcus, and the cerebellum. By contrast, LOIs with the highest participation rate in the top discriminative triplets were situated mainly in posterior regions such as the left cerebellum, the left superior parietal gyrus, the lateral occipital gyrus bilaterally, and the right angular gyrus.

We next explored the temporal properties of the top patterns. First, we examined the durations of the MRPs and the MDPs at the individual subject level. We extracted the duration for each top-ranked pattern by the time difference between the first and last event of the pattern. Following a similar procedure as for the spatial properties, we computed ten one-dimensional histograms for each subject, one for each cutoff value between the top $10^2$ and $10^6$ and pattern type. *Fig 7* depicts the duration pattern histograms for MRP and MDP types for seven subjects.

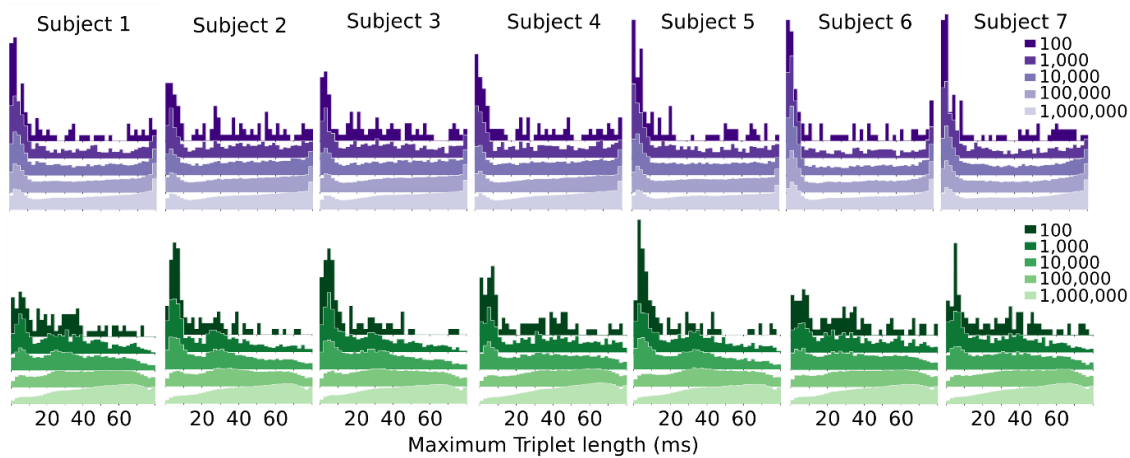

**Fig 7. Duration density histograms.** Each panel shows the density histogram of the patterns of durations. Colors refer to the cutoff value in the range of 102 to 106. The top and bottom panels show the histograms for the MRPs and MDPs, respectively. Each column depicts the results obtained for a single subject.

The figure shows that for both the MRPs and MDPs the characteristics of the timespan patterns changed as lower-ranked patterns were added to the analysis. Thus, we divided our observations into two separate ranges of cutoff values, $10^2$ to $10^3$, and $10^4$ to $10^6$. When inspecting the time span of $10^2$ to $10^3$ in the MDPs, two distinctive peaks emerged at ~7ms and ~35ms, consistently across subjects, whereas examining the same range of MRPs showed only a single peak at ~2ms. The inspection of MDPs in the range of $10^4$ to $10^6$ showed that the two peaks at ~7ms and ~35ms diminished. However, there was an increase in the range of ~10ms to ~75ms of the time spans with a less prominent peak at 80ms compared to the MRPs. Investigation of the most frequently repeating pattern histograms revealed that the peak of the short timespan at ~2ms diminished but was still present, but that a more prominent peak emerged at the ~80 ms time-span.

We then examined the temporal structure within each pattern. We conducted an analysis similar to the one used for the spatial relations. Each pattern classified within the top-ranked patterns was placed on coordinate in two-dimensional space. The first coordinate was the time delay between the first and second events within the pattern, and the second coordinate was the time delay between the second and third events within the pattern. Note that since only patterns with a duration of 80 milliseconds were inspected, the summation of the first and second delays could not exceed 80 milliseconds. We followed the same procedure as for the spatial relations analysis for computing ten two-dimensional histograms per subject, one for each pattern type and a cutoff value for the number of top patterns. To assess the statistical significance of the results, we conducted permutation tests as described in the *Methods* section. Since the results showed a consistent formation across subjects, for purposes of illustration, *Fig 8* depicts the median p-values across subjects obtained for each bin in the two-dimensional histograms.

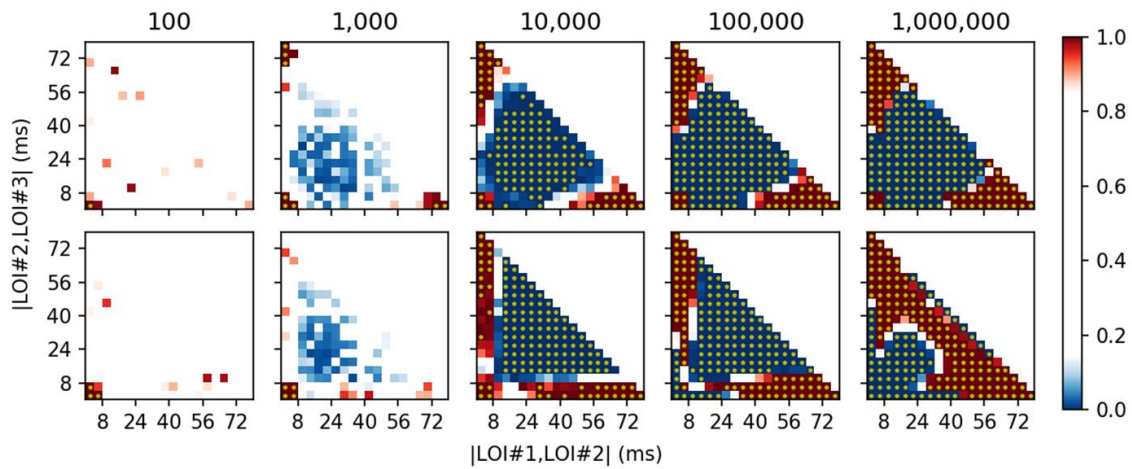

**Fig 8. Temporal relations between events in the top patterns.** Each panel shows the median p-values across subjects for temporal differences or delays. The delay between the first and second events and the delay between the second and third events are shown on the x-axis and y-axis, respectively, for each pattern found. The top and bottom rows show the histograms for the MRPs and MDPs, respectively. Each column shows the results obtained for a different cutoff value yielding a different number of patterns. Red and blue indicate greater or lesser counts in the data respectively compared to the permutation counts, and yellow asterisks depict bins with a $p-value < 0.001$ for both tails.

This analysis of the temporal relations between events within a pattern again revealed a symmetrical and organized formation for all pattern types. Examining the MRPs across the various cutoff values revealed a basic structure that became more prevalent as the cutoff values increased. This structure was comprised of three clusters in which more patterns were found and a complementary cluster where fewer patterns were found. The three clusters exceeding the cutoff corresponded to MRPs that tended to be composed of either simultaneous occurring events, or one pair of events occurring in close temporal proximity while the remaining event occurred at least 40ms later.

Exploring the temporal relations of the MDPs for cutoff values in the range of $10^2$ to $10^4$ revealed a similar organization of the top patterns. However, from a cutoff value of $10^4$ the differences between MRPs and MDPs became even greater as less discriminative patterns were added to the analysis. In the top $10^4$ MDPs, there were three clusters above cutoff: the simultaneously occurring events, and the combination of one short delay of up to 8ms, and a longer delay of at least 20ms. As the cutoff value became larger and less discriminative, patterns were added to the analysis, and the formation changed drastically. The simultaneously occurring cluster diminished until it disappeared, and instead, the two other clusters were associated and formed a single cluster so that most of the patterns were composed of two delays of at least ~28ms between pattern events. This drastic change in the organization of MDPs suggests that the characteristics of the top $10^4$ patterns differed significantly from less discriminative patterns, which also have non-random characteristics.

Based on our findings that MDPs have specific novel characteristics, we attempted to decode behavior at the single-trial level. However, a single-trial classification depends on the sparsity of the features. In other words, if a pattern is very discriminative but only occurs in five out of the ~180 trials, its contribution to classification is negligible. For that reason, we decided to test whether a relaxation of the temporal precision requirement could affect the classification results. For example, relaxation of the temporal precision by $\tau$ would mean that a pattern would be detected if a match was found with a temporal precision of $\tau * 2 + 1$ milliseconds. This relaxation of the requirement could allow a single feature to possibly affect many more test trials. The decoding procedure and the decoding evaluation are described in detail in the *Methods* section. *Fig 9* depicts the single-trial classification results obtained for each of the seven subjects shown in the panels, for a range of maximal pattern duration values (depicted by color) as a function of the temporal precision values.

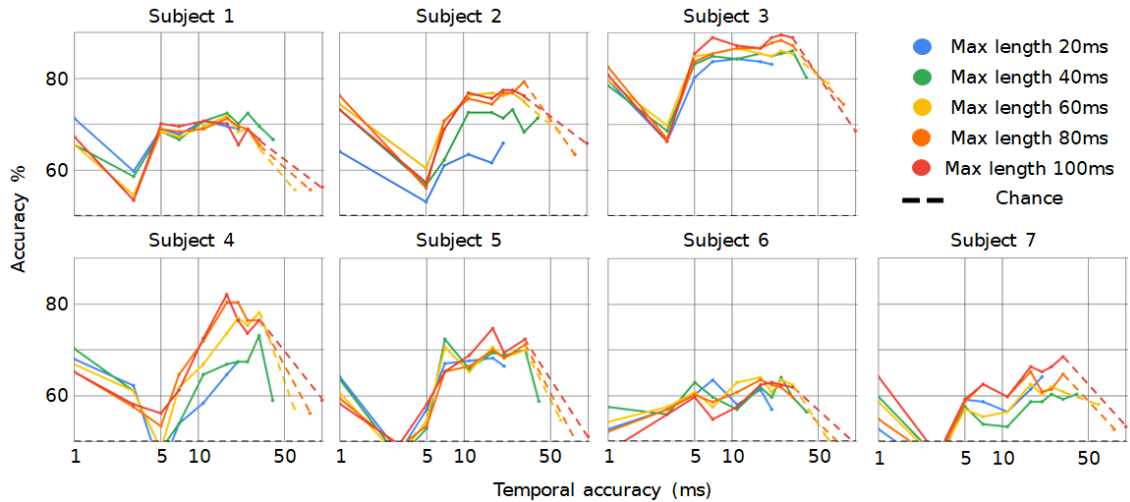

**Fig 9. Single-trial classification.** Each of the seven panels shows the classification results obtained for a single subject. Curves present the single-trial leave-one-out test-set accuracy as a function of temporal precision. Colors depict the maximum pattern time spans. The dashed black line indicates the chance level at 50%.

Investigating the patterns' temporal precision revealed several features. For all subjects, decoding based on the discriminative patterns as carried out in the naïve decoding procedure achieved results exceeding chance. However, there was considerable variability in decoding accuracies between subjects, ranging from ~95% accuracy for Subject 3 to only 66.7% for Subject 6. In all subjects, the best decoding results were obtained for a temporal precision in the range 17-31 milliseconds. In three of the seven subjects, there was a drop between 1 and 5 ms in decoding performance that may indicate that some patterns had very accurate timing. In all subjects and for all maximal length values, the decoder accuracy presented a prominent drop beyond the temporal resolution of 31ms. This drop highlights that assuming a temporal precision of 40-100ms by disruption of pattern timing impaired decoder performance profoundly.

Although most of the results showed high cohesiveness across subjects, the decoding accuracies varied broadly. We thus searched for a basic measure that could explain this variability. For each CCD we computed the variability around the time of mini-ER detection in a window of a hundred milliseconds. *Fig 10* depicts these variabilities for all subjects, where each line shows the variability in a single LOI. Inspection of the CCD variability at the time of mini-ER detection revealed that some LOIs exhibited higher variability than others. For each subject we tested for correlations with decoding accuracy. *Fig 10* bottom right panel in *Fig 10* shows the relationship between this measure of variability of variances and mean decoding accuracy. Spearman's rho correlation was used to assess the relationship between the measure of variability of variabilities and the mean decoding accuracy across maximal window lengths at a temporal accuracy of 31ms. There was a significant correlation between the two ($p = 0.03, r = -0.82$).

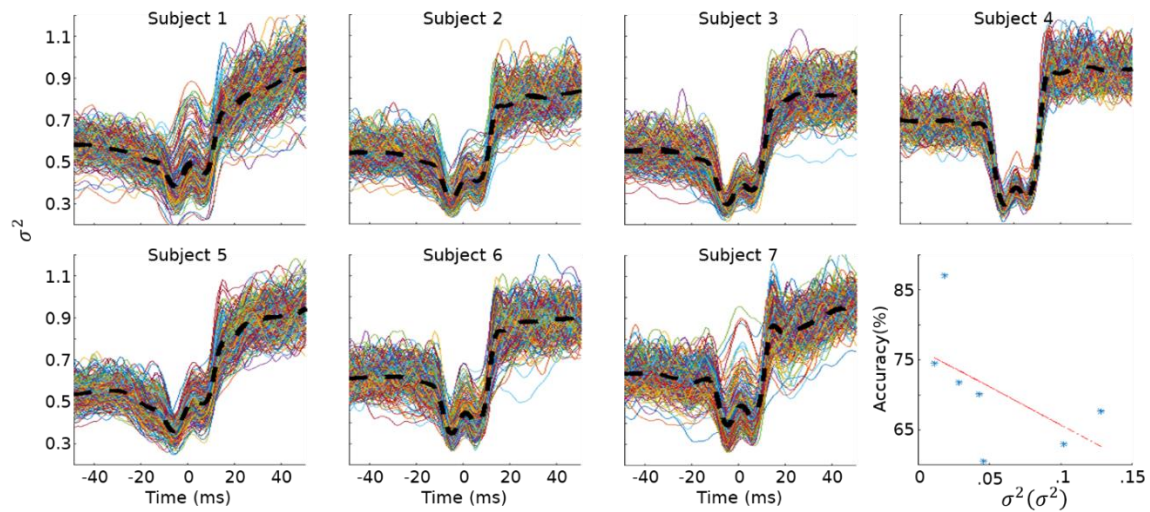

**Fig 10. Variances around mini-ER detection times and correlation to decoding accuracy.** Each of the seven panels shows the variability results obtained for a single subject. In each panel, each trace shows the variability in a single LOI for a window of a hundred milliseconds around the times of mini-ER detections. The dashed black line indicates the mean of all traces in the panel. The bottom right panel shows the relationship between the mean decoding accuracy and the variance of LOI variances. The mean decoding accuracy was computed across maximum window lengths for a temporal resolution of 31ms. Each point shows the two values for a single subject.

## Discussion

This study constituted a preliminary effort to investigate the properties of brain-wide spatio-temporal patterns. We detected local transient events and their global organization as spatio-temporal patterns in the data of subjects engaged in a sensorimotor synchronization task. We focused our analysis on two types of patterns, the *most frequently repeating patterns* and the *most discriminative patterns,* to concentrate on patterns that were the most pertinent to the subjects' behavior out of the initial ~50,000,000 unique patterns. Based on these two types of patterns, we investigated the prevalence across patterns of properties such as the duration of patterns, the spatial and temporal relations between pattern sub-units and the patterns' temporal precision. We found that patterns exhibited a typical formation in both space and time across subjects. Although most of these characteristics were common to both the discriminative and most frequently repeating patterns, there were several significant differences between them. A simplistic single-trial decoding approach based on the discriminative patterns indicated a mean accuracy of ~74% across subjects and revealed a resemblance between patterns and the temporal precision of the neural assemblies.

We assessed the temporal precision of the discriminative patterns and their specificity to the behavioral conditions using a single-trial decoding approach. We found that in all subjects, a peak in decoding accuracy occurred for a temporal resolution between **17**ms and 31ms. This range of temporal precisions may point to a grouping or sampling mechanism based on a single cycle of gamma oscillation, which is consistent with the hypothesis and findings in the literature suggesting that neural assemblies synchronize their firing within a gamma cycle to accumulate a sufficient number of neurons to actuate downstream populations (Buzsáki, 2010; Hahn et al., 2014). The temporal precision we found for global activation patterns is similar to the temporal resolution of 10-30ms observed in cell assemblies found locally in the hippocampus (Harris et al., 2003).

Purely in terms of sample size one would expect that relaxing the temporal precision requirements from 1 to 5ms would reduce feature sparsity, and that the performance of the decoders would improve. However, in three out of the seven subjects, there was a drop between 1 and 5 ms in decoding accuracy. This type of temporal precision has been found locally in the visual cortex (Usrey and Reid, 1999), the prefrontal cortex (Constantinidis et al., 2001) and in the hippocampus (Csicsvari et al., 1998). On a global scale, this result, together with the findings for the patterns' temporal and spatial relations, lends weight to the supposition that remote sources show highly precise timings of less than 5ms, which is consistent with our previous studies (Felsenstein et al., 2019b; Tal and Abeles, 2016). They raise the question of which models and theories of network activity can account for such a high temporal precision between remote sources. One possible explanation could be a partial overlap between synfire chains (Abeles et al., 2004; Hayon et al., 2005) that may be implemented through additional corticocortical or thalamocortical connections (Asai et al., 2008; Frostig et al., 2008; Mohajerani et al., 2013).

Note that in all subjects, relaxation of the temporal precision criteria beyond 31ms considerably reduced pattern specificity to behavior, which underscores the need to use neuroimaging methods with a suitable temporal resolution. Based on our findings that sources at large distances are involved in discriminative patterns, the ability to measure relevant signals in remote brain regions, in conjunction with this temporal precision requirement, is crucial to further study the possibility of global packet-like cortical communication.

This study contributes to the growing number of works reporting on the presence, characteristics, and significance of transient events to behavior. The findings here are congruent with previous studies showing that spatiotemporal mini-ER patterns show greater precision than expected by chance (Tal and Abeles, 2016) and that delays between events in remote brain regions can decode subjects' behavior at the single-trial level (Felsenstein et al., 2019b, and Study II of this dissertation). In the present study, we showed that sequences of transient events have a typical spatial and temporal organization in relation to the subject's behavior that is consistent across subjects. Not only is the rate of events in various sources predictive of subjects' behavior (Abeles, 2014; Shin et al., 2017a), but the spatiotemporal patterns of these events across the brain can predict subjects' behavior significantly beyond chance

Each discriminative pattern was detected in only 5% of the trials on average. If some patterns are specific to a subject's behavior, what accounts for the fact that these patterns do not repeat each time the same behavior takes place and, as a result, exhibit higher decoding performance? The explanation is both conceptual and technical.

Conceptually, we first need to consider whether each time a subject performs a task, the same cognitive process occurs. It is plausible that subjects' prior experience on previous trials or spontaneous activity leads to different routes and patterns enabling a subject to execute the same action or decision-making process. From a decoding perspective the fact that a pattern repeats multiple times does not guarantee that it relates to the same mental process. The MEG spatial resolution is about a centimeter (Sharon et al., 2007; Vorwerk et al., 2014), and the human cortex contains ~50,000,000 neurons in a cubic centimeter (Abeles, 1991). It is possible that activations of various subgroups of neurons or cell assemblies might render the same signal as an mini-ER. If so, although the patterns of cell assemblies are highly specific to subjects' behavior, the patterns of mini-ER we detected are not as specific.

The low number of repetitions or the lesser specificity of a behavioral condition of a pattern could also originate from noise in the detection of mini-ERs due to the detection threshold value used in the mini-ER detection procedure. An overly lax threshold would result in false detection of events that could fail to detect pattern specificity to behavior, and an overly high threshold would result in fewer repetitions identified for each pattern. When we investigated the high variability across subjects' decoding results, we hypothesized that inaccurate detections of mini-ERs might hinder the decoding process. This hypothesis was supported by the correlation between decoder accuracy and variability in the mini-ERs. Future works should consider a more sophisticated methodology to set the detection thresholds for the mini-ERs. One possible mechanism would be to define differential thresholds for each trial and source, as was done when detecting beta-events (Sherman et al., 2016; Shin et al., 2017a). A second possible method would be to look for phase locking of mini-ERs to oscillatory activity such as theta-oscillations in the case of cell assemblies (Harris et al., 2003; Sirota et al., 2008).

Although the findings provide initial support for the claim that the patterns found in this study are a global realization of cell assemblies theory, we do not have a full understanding of the neurophysiological basis of mini-ERs. If future LFP studies can link mini-ERs with cell assembly activations, this will lend additional weight to the supposition that the theory of cell assemblies could be a part of a nested principle. Unlike the case of cell assemblies, the measurements obtained by the MEG suffer from a lower resolution in space, and whereas our method was designed to handle temporal imprecisions. Thus, future studies could extend this method to allow for spatial imprecisions. Finally, in the current study, discriminative patterns were analyzed as a whole. Future studies should examine a specific behavioral condition that could reveal more precise and homogenous characteristics.

This study provides an initial exploration of the spatial and temporal properties of brain-wide patterns composed of transient cortical events. Our results are the first, to the best of our knowledge, to demonstrate the existence of brain-wide packet-based sequential communication in humans through the cortex and cerebellum. We found that behavioral-specific patterns are distributed spatially and that remote brain regions can manifest highly precise timings of few milliseconds. The patterns identified in this study show cohesive asymmetrical spatial relations between pattern sources, which call for further research. Surprisingly, we found that the temporal resolution between sources within the global patterns was similar to the temporal resolution found within a brain region such as the hippocampus (Harris et al., 2003). Overall these results underscore the importance of measuring neural activity across distant brain locations with high temporal resolution.